# The Web Graph of a Tourism System

Rodolfo Baggio

Master in Economics and Tourism, Bocconi University, Milan, Italy

School of Tourism, The University of Queensland, Australia

**Present address:**

Master in Economics and Tourism - Bocconi University

via Sarfatti, 25

20136 Milan, Italy

Phone: +39 02 5836 5437

Fax: +39 02 5836 5439

Email: rodolfo.baggio@unibocconi.it

**Abstract**

The website network of a tourism destination is examined. Network theoretic metrics are used to gauge the static and dynamic characteristics of the webspace. The topology of the network is found partly similar to the one exhibited by similar systems. However, some differences are found, mainly due to the relatively poor connectivity and clusterisation of the network. These results are interpreted by considering the formation mechanisms and the connotation of the linkages between websites. Clustering and assortativity coefficients are proposed as quantitative estimations of the degree of collaboration and cooperation among destination stakeholders.

# 1. Introduction

The vast catalogue of studies on complex networks which has been compiled in the last few years is missing an important component: the tourism industry (see the reviews in: [1] [2] [3] [4]). This paper aims at filling this gap and presenting an investigation on the websites network of a tourism destination.

In the second part of last century, tourism has become probably the largest economic sector of the World economy. In broad terms, according to the last estimates of the World Travel and Tourism Council [5], it is expected to total 10.6% of GDP and more than 200 million jobs. And the growth is thought to continue for the next years at a rate close to 5% per year.

The boundaries of the tourism and travel industry are fairly indefinite. It brings together, and strongly influences, segments from a number of different activities with a wide variety of products and services exhibiting very little homogeneity. Moreover, in the last decades tourism has become an extremely dynamic system [6]. There is probably no other economic sector with such a diversity and this has raised the question of whether tourism and travel should even be classified as an industry by itself, in the traditional sense of manufacturing or trade.

In the last years, the globalisation enabled by technology development and by less expensive travel costs has greatly increased competition. The intensified marketing efforts of all tourism organisations has led to a more effective approach: the destination management approach [7].

The spectrum of definitions describing a destination is extremely broad, and there are many difficulties in setting clear boundaries to a Tourism Destination (TD). In general, every place for a holiday, every place to visit may be considered a destination. In broad terms, a tourism destination may be intended as a geographical area that offers the tourist the opportunity of exploiting a variety of attractions and services [8].

The supply is provided by a more or less definite set of private and public organisations and companies that, in the ideal case, collaborate and coordinate their efforts in order to maximise their profits and to assure a balanced and sustainable progress of the local resources, avoiding any possible threat to people and environment [9]. Destinations exist at a number of different interrelated geographical levels, so that we may envision a hierarchy of TDs forming a global tourism system.[10].

Apart from the definition problems, a TD, essentially a socio-economic system, is the archetype of a complex adaptive system (CAS). It shares many (if not all) of the characteristics usually associated with a CAS: non-linear relationships among the system components (private and public companies and associations), self-organisation of the structures, robustness to external shocks [11] [12] [13] [14]. In recent times, several authors have approached the study of TDs by using the perspective of complex systems science [15] [6] [16], and a number of symptoms has been visibly identified both from a qualitative and a quantitative viewpoint. For example, it has been shown that Zipf-like relationships exist in the distribution of arrivals of tourists at destinations [17], both at a country level and within countries [18]. Moreover, the analysis of time series of tourism data (arrivals or night visits) has allowed to highlight a noticeable robustness (resilience) of several TDs to external or internal shocks such as epidemics [19], international crises [20], political instabilities [21] or conflicts [22]. These effects have also been highlighted when considering a large TD along with one of its subsystems [18]. This indication of self-organisation and self-similarity are a clear signature of nonlinear dynamic processes typical of many complex systems [23].

The intangibility of a tourism product stresses its information component so that it is always described as an "information intensive" one [24]. Thus, it is not surprising that the relationship between tourism and information technology is very strict. This sort of genetic tie originates at the dawn of the electronic computer history, at the end of the 1960s with the deployment of the first computerised reservation systems. The evolution of the two industries has almost always been parallel and, not unexpectedly, nowadays the Internet has in travel and tourism organisations its most numerous and important component.

The Internet age has allowed the development of new ways for producing and distributing travel and tourism services. Web-based approaches and technologies are helping suppliers and agencies in reducing service costs and attracting customers [9]. A website looks to be a major (and, probably, it will be the only one in the future) tool to conduct business in the tourism field [25]. According to PhoCusWright's estimates [26], for example, online sales in the U.S., 35% of the travel market in 2005, will account for more than 50% in 2006.

## 2. The web space of a tourism destination

The websites of a tourism destination have been analysed. The destination is the island of Elba, off the coast of Tuscany, Italy, in the heart of the western Mediterranean Sea. It is an important environmental resource; its geographic position, temperate climate, and the variety and beauty of its landscapes, coast and sea, make it a renown tourist destination.

As for many other destinations, the Web has become, in the last years, an important means of promotion and of commercialisation for the whole community of Elban tourism operators and its diffusion has reached almost the whole population [27] [28].

The elements of the network examined are the websites belonging to the core tourism operators: accommodation (hotels, residences, camping sites etc.), intermediaries (travel agencies and tour operators), transport, regulation bodies, services. The whole size of the network is not huge, it comprises 468 elements. This size can be thought sufficient to show statistical properties in a meaningful way (see the discussion in: [29] and [30]).

The websites have been analysed considering them as the nodes of a complex network. Links among the websites have been counted by using a simple crawler and complementing the data obtained with a visual inspection of the websites. Besides that, links connecting the Elban websites to the rest of the Web, in both directions, have been identified. In what follows E denotes the graph defined by the edges connecting Elban websites (i.e. links connecting exclusively destination stakeholders located at Elba), W is the set of links between Elban websites and the rest of the World Wide Web. All links are considered of directed nature. Figure 1 (drawn with Pajek [31]) gives a graphical representation of the E network thus obtained.

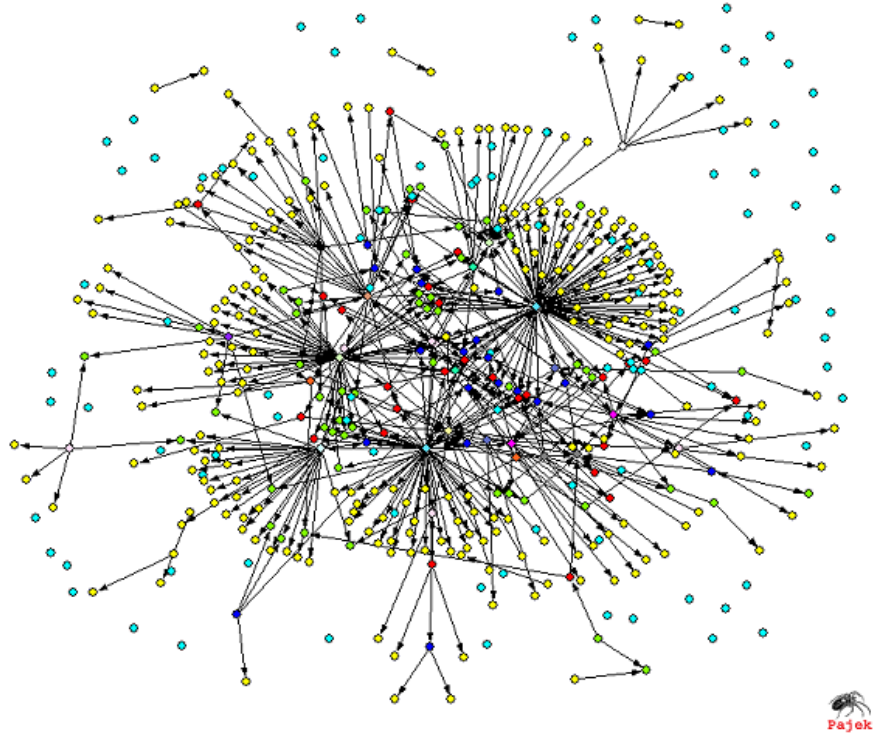

Figure 1 : The network of the Elban tourism websites (drawn with Pajek [31]; colour online)

3. **The results of the statistical analysis**

The E network is rather sparse, its link density is d = 0.002 and almost 21% of the websites have no connection whatsoever with other sites. The diameter is D = 11, the average distance L = 4.5 and the global clustering coefficient C = 0.003.

Key parameters characterising the structure of a directed network are the in-degree ($k_{in}$) and out-degree ($k_{out}$) distributions. Both the E and W networks, as portrayed in fig. 2 and 3, display an almost perfect power law decay $P(k) \sim k^{-\gamma}$. The cumulative degree distributions are shown in figure 2 and 3. The exponents calculated for the networks are listed in table 1.

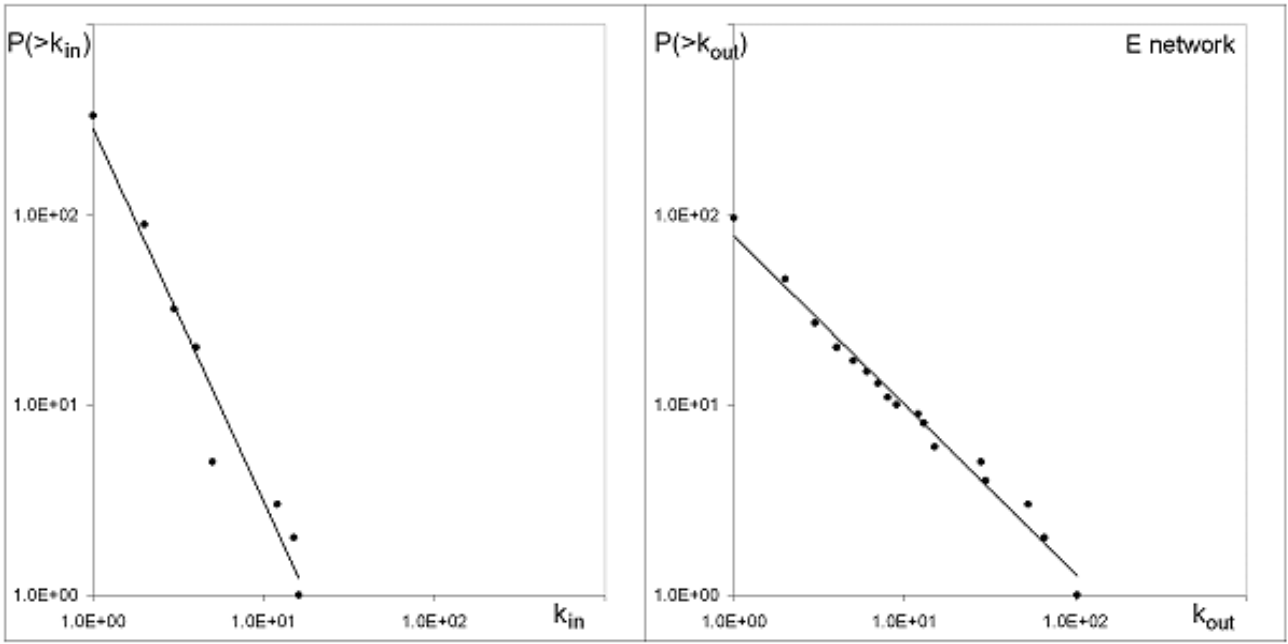

Figure 2: Cumulative in-degree ($k_{in}$) and out-degree ($k_{out}$) distributions for the E network of the tourism websites of the Elba island

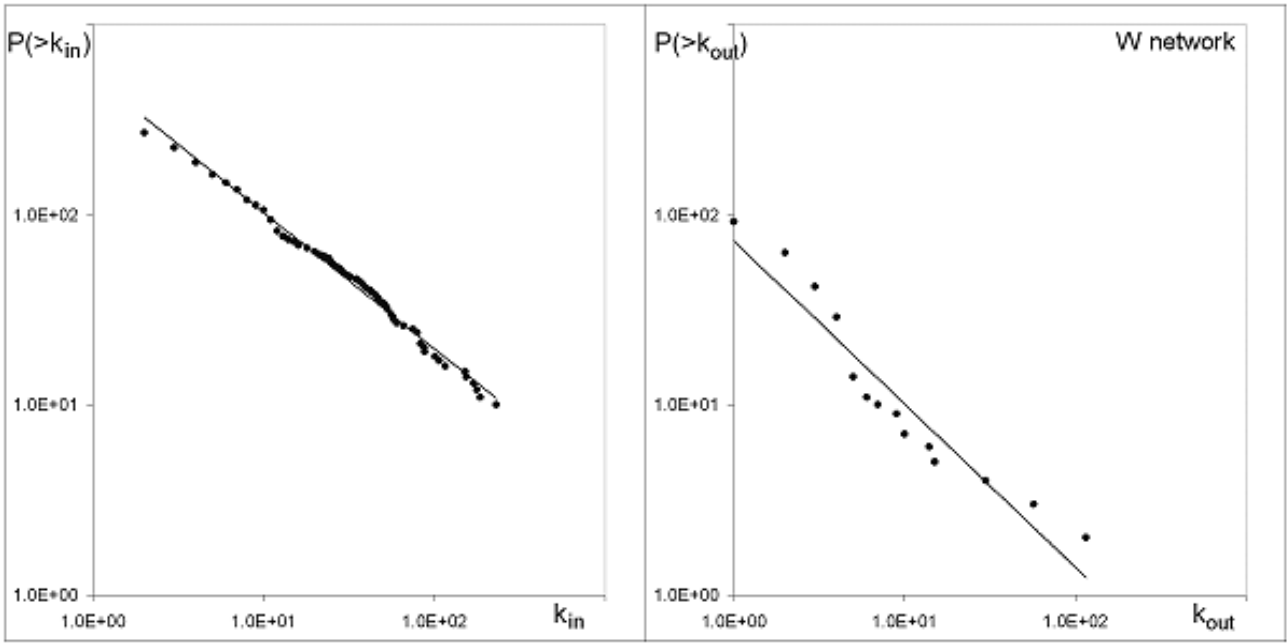

Figure 3: Cumulative in-degree ($k_{in}$) and out-degree ($k_{out}$) distributions for the W network of the tourism websites of the Elba island

Table 1. Degree distribution exponents

| Network | Degree distribution exponent | |
|---|---|---|
| E | $\gamma_{out}$ | 1.89 |
|   | $\gamma_{in}$ | 2.96 |
| W | $\gamma_{out}$ | 1.86 |
|   | $\gamma_{in}$ | 1.72 |

All the exponents (with the exception of $\gamma_{in}$ for the E network) are lower than those typically measured for the Web ($\gamma_{in} \sim 2.1$ and $\gamma_{out} \sim 2.7$ [32]) showing thus a more skewed and "sparse" distribution which may be seen as a very low propensity to reference the external world.

A spectral analysis confirms the main topological characteristics of the E network.

The shape of the spectral density $\rho(\kappa)$ of a graph is known to be an indicator of the topological properties of a network [33] [34]. For random graphs with a giant connected component it converges to a semicircle following Wigner's law [35]. All other cases see different distributions: a highly skewed multi-peaked structure for a small-world network and a triangular shape for scale-free graphs. As figure 4a shows, the power law behaviour of the degree distribution for the E network is evident.

The $\rho(\lambda)$ spectral distribution of the Laplacian matrix associated to the E network is shown in Figure 4b. It can be noted that a high number of the Laplacian eigenvalues is null. This is an indication [36] of the scarce connectedness of the network. The multiplicity of the null eigenvalue, in fact, corresponds to the number of the connected components of the network.

The general topological properties of the World Wide Web have been studied by a number of authors. In particular, it has been possible to highlight a complex structure of its components (web pages or websites). According to Broder et al. [37], the structure has a bow-tie shape, in which it is possible to recognise a number of components characterised by their connectivity characteristics. The model, widely accepted, sees a strongly connected component (SCC), formed by all pages mutually connected by a directed link, an in-component (IN) and an out-component (OUT), formed by nodes connected to the SCC in a unidirectional way plus a series of secondary structures such as TENDRILS, containing pages that cannot reach the SCC and cannot be reached from it, TUBES, directly linking the IN and OUT parts without crossing the SCC and some disconnected elements (DCC), similar to isolated islands, with no connection at all to the other components.

This structure has also been identified in several sub-networks of the whole Web, presupposing a self-similar configuration for the Web [38].

The Elban network (E network), besides the general low connectivity among its websites, still exhibits a bow-tie structure. Table 2 displays the estimated proportions for the bow-tie components along with the values accepted for the whole Web.

Table 2 Relative size of the components for the E network and the Web ([17] [22]) in the hypothesis of a bow-tie structure

|  | E Network (%) | Web (%) |
|---|---|---|
| **SCC** | 3.4 | 28.0 |
| **IN** | 2.1 | 21.0 |
| **OUT** | 52.4 | 21.0 |
| **TENDRILS** | 15.6 | 21.0 |
| **TUBE** | 1.3 | 9.0 |
| **DCC** | 25.2 | |

Local and global efficiency [39] are: $E_{loc}$ = 0.0145 and $E_{glob}$ = 0.16981. These values are sensibly lower than those found for similar systems. The $E_{loc}$ value is consistent with the low clustering coefficient seen above.

**4. Discussion and interpretation of the results**

The studies conducted so far on the structure of the WWW have claimed a substantial self-similarity. However, some discrepancies have been found when considering certain category-specific groupings of websites [40] which show a significant deviation from a power law in the degree distribution.

The data presented in the previous section give a picture that may look contradictory. While some values (diameter, average distance or power law scaling in the degree distribution) are consistent with those generally found in WWW analyses, others exhibit significant differences. The general connectivity is quite lower, the distribution of the degrees is more skewed, and the overall structure of the Elban web graph, as the bow-tie data shows, strongly departs form the generally accepted values. It should be concluded that a strict self-similarity cannot be called for.

An interpretation of these results must take into account the formation mechanism of the connections among websites and the connotation of the network analysed. Not having sufficient data to trace an evolutionary history of the TD and its web space, it would be possible to model this evolution by using one of the many theoretical models proposed in the literature [4]. In the present case, however, a different and wider reading of our results can be attempted, which takes into account the nature of the network under investigation.

In developed tourism areas like Elba, where the Web has a very high diffusion and plays a crucial role as a means for communicating or conducting businesses [27] [28], the network of websites denotes more than just an artificial technological network; the web space of a tourism destination can be seen as a close representation of the underlying economic and social network. The structure of hyperlinks form patterns based on the plans and the designs of individuals or organisations owning the websites. A growing literature suggests that these networks directly reflect offline relations among social actors and support specific social or communicative functions [41] [42] [43]. This relationship between cyberspace and the physical world is reciprocal: on one side, the online linkages represent and complement social relations in the offline world; on the other side, offline interactions can influence the way in which online relationships are established and developed [44] [45]. In this respect, the layout of such a network can be seen as an expression of the characteristics of the structure of the socio-economic system from which it originates.

Under this assumption, the general low connectivity and low clustering characteristics of the Elban web graph is a clear indication of very limited degree of collaboration among the components of the TD. The capability of achieving an effective cooperation among the various elements of the network is also challenged by the very low efficiency (both local and global) of the network, in the common interpretation of efficiency as a measure of how well information is exchanged over the network [39].

A confirmation of this understanding comes from previous qualitative studies on Elba TD [27] [28], which have argued that a low propensity to connect to the external world exist and it finds a reason in the strong independent way to conduct small family-run enterprises (the vast majority of the tourism businesses on the island). The higher than expected sparseness of the degree distribution of the W network, the set of connections between Elban websites and the rest of the Web, further confirms this reading, showing again a very low propensity to exchange with the "outside" world and the tourism system hierarchy. More work and larger samples of TDs are obviously needed before being able to confirm this conjecture and to ascertain the effective linkages among the components of the global system.

One more consideration can be made. Thinking of a web space as a virtual counterpart of an important economic and social system, an interesting quantity to measure is the assortativity coefficient, the extent of correlation existing between nodal degrees. Many other social networks studied so far are characterised by a positive assortative mixing, i.e. the edges preferably connect vertices with similar degrees. This is usually interpreted as a sign of collaboration among the actors of the network [3] [46] [47] [48]. The coefficient can be calculated as the Pearson correlation coefficient between the degrees of adjacent vertices in the network [47]. For the Elban network (E network) the coefficient is $r = -0.101 \pm 0.094$ (standard error is computed as in [47]). In spite of what it is usually found for a social network [3] [46] [47] [48], our coefficient is negative (although very small), meaning a certain reluctance of the components of our network to team up.

## 5. A proposal

In the scenario delineated in the previous sections, two of the quantities measured assume an important meaning. The clustering coefficient and the assortativity index can be used as quantitative assessments of the extent to which the tourism organisations work together collaborating or cooperating, i.e.: forming cohesive communities inside the destination. For their nature, the clustering coefficient can be thought of a *static* measurement, while the assortativity coefficient can be interpreted as expressing the tendency to form such communities.

The fragmented nature of the tourism industry, for the diversity of activities and organisations involved, is one of its natural traits, and the lack of cooperation among the actors of these networks is equally well known. This fragmented nature is considered a main reason for the need for cooperation. Many authors [49] [50] [51] [52] claim that a tourism system can have a balanced and planned evolution only through a process of shared information and decision making with all the stakeholders involved. This much sought-after capacity to work together is considered to be a crucial element for the success of a destination [53].

So far, collaboration and cooperation characteristics have been assessed by using qualitative methods (surveys, focus groups etc.) [54] [55]. Tourism managers and planners may now have at their disposal a quantitative measure, relatively easy to estimate, to gauge this phenomenon or to complement (and validate) their explorations.

## 6. Summary

The network formed by the websites of a tourism destination has been analysed, as part of a larger project on the structure of the relationships existing among the stakeholders of such systems. The statistical mechanics tools developed in the last years for this purpose [4] have been used to derive the main topological characteristics of this network.

Some of the results show a general agreement with similar results [37] [38] [56] [57], obtained by studying the Web and websites configurations. This may partially reinforce the idea of a substantial self-similarity in the structure of the Web space [38]. Some of the values, though, show different characteristics: basically a lower connectivity and a higher sparseness.

It is difficult to explain these differences as the history of the system's evolution is not known.

Nevertheless, if we accept the idea that, given the wide-reaching diffusion of the Internet, a web network is a close representation of the social network formed by the websites' owners [45], and we interpret the results presented here by adopting this viewpoint, then we see a faithful representation of the relationships among the actors of a typical tourism destination.

The outcomes presented here show how the statistical analysis of networks can render quite faithfully the structure of a peculiar social network. Moreover, it is proposed that clustering and assortativity coefficients be used to measure, quantitatively, the extent of collaboration or cooperation among the stakeholders in a tourism destination.

This is, at author's knowledge, one of the very first attempts to use these techniques in this field. Further work is under way to explain the dynamical evolution of a tourism system and to relate it with the methods traditionally used for its analysis. Furthermore, the application of simulation algorithms can suggest modifications to the structure of the network in order to optimise its features and behaviours.

A final consideration is in order. The issue of "legitimation" of the field of tourism studies as a discipline is an ongoing debate [58] [6] [59] [60]. Emphasis has been given, very often, to the necessity of more rigorous theoretical approaches. The author hopes that a wider diffusion of network analysis methods, mainly if they are considered as framed into the larger field known as *complexity science* [12], can provide, at least partly, a means to "reconceptualise" tourism studies [6].

## Acknowledgments

The author wishes to thank Chris Cooper, Noel Scott and Magda Antonioli Corigliano for the helpful discussions during the preparation of the manuscript. The author expresses also his gratitude for the useful suggestions and comments of an anonymous referee.

# 7. References

[1] R. Albert, and A.-L. Barabási, Rev. Mod. Phys. **74**, 47 (2002).
[2] S. N. Dorogovtsev, and J. F. F. Mendes, Adv. Phys. **51**, 1079 (2002).
[3] M. E. J. Newman, SIAM Rev. **45**, 167 (2003).
[4] S. Boccaletti *et al.*, Phys. Rep. **424**, 175 (2006).
[5] WTTC, *The 2005 Travel & Tourism Economic Research* (World Travel & Tourism Council, London, 2005).
[6] B. H. Farrell, and L. Twining-Ward, Annals of Tourism Research **31**, 274 (2004).
[7] J. R. B. Ritchie, and G. I. Crouch, *The Competitive Destination: A Sustainable Tourism Perspective* (CABI Publishing, Oxon, UK, 2003).
[8] J. Jafari, *Encyclopedia of Tourism* (Routledge, London, 2000).
[9] D. Buhalis, *eTourism: Information technology for strategic tourism management* (Pearson/Prentice-Hall, Harlow, UK, 2003).
[10] S. Wahab, and C. Cooper, *Tourism in the age of globalisation* (Routledge, New York, 2001).
[11] J. H. Holland, *Hidden Order: How Adaptation Builds Complexity* (Helix Books, Reading, MA, 1995).
[12] L. A. N. Amaral, and J. M. Ottino, Eur. Phys. J. B **38**, 147 (2004).
[13] W. B. Arthur, S. Durlauf, and D. Lane, *The Economy as an Evolving Complex System II* (Addison-Wesley, Reading, MA, 1997).
[14] Y. Bar-Yam, *Dynamics of Complex Systems* (Addison-Wesley, Reading, MA, 1997).
[15] B. Faulkner, and R. Russell, Pacific Tourism Review **1**, 93 (1997).
[16] B. McKercher, Tourism Management **20**, 425 (1999).
[17] M. H. Ulubaşoğlu, and B. R. Hazari, Journal of Economic Geography **4**, 459 (2004).
[18] R. Baggio, arXiv/physics/0701063 (2007).
[19] O. Dombey, Journal of Vacation Marketing **10**, 4 (2003).
[20] J. Eugenio-Martin, M. T. Sinclair, and I. Yeoman, Journal of Travel and Tourism Marketing **19** 23 (2005).
[21] P. K. Narayan, Tourism Economics **11**, 351 (2005).
[22] H. Aly, and M. C. Strazicich, (Economic Research Forum, 2002).
[23] T. Komulainen, in *Complex Systems - Science on the Edge of Chaos (Report 145, October 2004)*, edited by H. Hyötyniemi (Helsinki University of Technology, Control Engineering Laboratory, Helsinki, 2004).
[24] A. Poon, *Tourism, technology and competitive strategies* (CAB International, Oxon, UK, 1993).
[25] R. Law, K. Leung, and J. Wong, International Journal of Contemporary Hospitality Management **16**, 100 (2004 ).
[26] PhoCusWright, *Online Travel Overview: Market Size and Forecasts 2004-2006* (PhoCusWright, Sherman, CT, 2005).
[27] H. Pechlaner *et al.*, in *Information and Communication Technologies in Tourism*, edited by A. J. Frew, M. Hitz, and P. O'Connor (Springer, Wien, 2003), pp. 105.
[28] V. Tallinucci, and M. Testa, *Marketing per le isole* (Franco Angeli, Milano 2006).
[29] P. Angeloudis, and D. Fisk, Physica A **367** 553 (2006).
[30] J. A. Dunne, R. J. Williams, and N. D. Martinez, Proc. Natl. Acad. Sci. USA **99**, 12917 (2002).
[31] W. de Nooy, A. Mrvar, and V. Batagelj, *Exploratory Social Network Analysis with Pajek* (Cambridge University Press, Cambridge, 2005).
[32] R. Pastor-Satorras, and A. Vespignani, *Evolution and structure of the Internet - A Statistical Physics Approach* (Cambridge University Press, Cambridge, UK, 2004).

[33] I. J. Farkas *et al.*, Phys. Rev. E **64**, 026704 (2001).
[34] K.-I. Goh, B. Kahn, and D. Kim, Phys. Rev. E **64**, 051903 (2001).
[35] E. P. Wigner, Ann. Math. **67**, 325 (1958).
[36] B. Mohar, in *Graph Theory, Combinatorics, and Applications*, edited by Y. Alavi *et al.* (Wiley, New York, 1991), pp. 871.
[37] A. Z. Broder *et al.*, Computer Networks **33**, 309 (2000).
[38] S. Dill *et al.*, ACM TOIT **2**, 205 (2002).
[39] V. Latora, and M. Marchiori, Phys. Rev. Lett. **87**, 198701 (2001).
[40] D. M. Pennock *et al.*, Proc. Natl. Acad. Sci. USA **99**, 5207 (2002).
[41] M. H. Jackson, in *Journal of Computer-Mediated Communication [On-line]* 1997).
[42] H. W. Park, Connections **25**, 49 (2003).
[43] H. W. Park, and M. Thelwall, in *Journal of Computer Mediated Communication [On-line]*2003 ).
[44] S. A. Birnie, and P. Horvath, in *Journal of Computer-Mediated Communication [On-line]* 2002).
[45] B. Wellman, Science **293** 2031 (2001 ).
[46] M. Catanzaro, G. Caldarelli, and L. Pietronero, Phys. Rev. E **70**, 037101 (2004).
[47] M. E. J. Newman, Phys. Rev. E **67**, 026126 (2002).
[48] M. E. J. Newman, Phys. Rev. Lett. **89**, 208701 (2002).
[49] D. J. Timothy, Journal of Sustainable Tourism **6**, 52 (1998).
[50] L. Roberts, and F. Simpson, Journal of Sustainable Tourism **7**, 314 (1999).
[51] C. Huxham, Public Management Review **5**, 401 (2003 ).
[52] T. B. Jamal, and D. Getz, Annals of Tourism Research **22**, 186 (1995).
[53] B. Bramwell, and B. Lane, *Tourism Collaboration and Partnerships: Politics Practice and Sustainability* (Channel View Publications, Clevedon, UK, 2000).
[54] T. B. Jamal, S. Stein, and T. Harper, Journal of Planning Education and Research **22**, 164 (2002).
[55] A. Ladkin, and A. M. Bertramini, Current Issues in Tourism **5**, 71 (2002).
[56] J. M. Kleinberg *et al.*, in *Lecture Notes in Computer Science*, edited by T. Asano *et al.* (Springer, Berlin, 1999), pp. 1.
[57] R. Kumar *et al.*, in 9th ACM SIGMOD-SIGACT-SIGART symposium on Principles of database systems Dallas, TX 2000), pp. 1.
[58] C. M. Echtner, and T. B. Jamal, Annals of Tourism Research **24**, 868 (1997).
[59] N. Leiper, Annals of Tourism Research **27**, 805 (2000).
[60] J. Tribe, Annals of Tourism Research **24**, 638 (1997).